\documentclass[11pt,a4paper,english,nofootinbib,,superscriptaddress]{revtex4}
\usepackage{lmodern}

\usepackage[T1]{fontenc}
\usepackage[latin9]{inputenc}
\usepackage{babel}

\usepackage{amsmath}
\usepackage{graphicx}
\usepackage{amssymb}
\usepackage{esint}
\usepackage[unicode=true, pdfusetitle,
 bookmarks=true,bookmarksnumbered=false,bookmarksopen=false,
 breaklinks=false,pdfborder={0 0 1},backref=false,colorlinks=false]
 {hyperref}
\def\b{\begin{equation}}
\def\e{\end{equation}}

\makeatletter
\@ifundefined{textcolor}{}
{%
 \definecolor{BLACK}{gray}{0}
 \definecolor{WHITE}{gray}{1}
 \definecolor{RED}{rgb}{1,0,0}
 \definecolor{GREEN}{rgb}{0,1,0}
 \definecolor{BLUE}{rgb}{0,0,1}
 \definecolor{CYAN}{cmyk}{1,0,0,0}
 \definecolor{MAGENTA}{cmyk}{0,1,0,0}
 \definecolor{YELLOW}{cmyk}{0,0,1,0}
 }

\usepackage{latexsym}\usepackage{bm}

\makeatother

\makeatother

\begin{document}

\title{Weyl-Invariant Higher Curvature Gravity Theories in $n$ Dimensions}

\author{M. Reza Tanhayi}

\email{m_tanhayi@iauctb.ac.ir}

\affiliation{Department of Physics,\\
 Middle East Technical University, 06531, Ankara, Turkey}

\affiliation{Department of Physics,\\
Islamic Azad University Central Tehran Branch, Tehran, Iran}

\author{Suat Dengiz}

\email{suat.dengiz@metu.edu.tr}

\affiliation{Department of Physics,\\
 Middle East Technical University, 06531, Ankara, Turkey}

\author{Bayram Tekin}

\email{btekin@metu.edu.tr}

\affiliation{Department of Physics,\\
 Middle East Technical University, 06531, Ankara, Turkey}

\date{\today}
\begin{abstract}

We study the particle spectrum and the unitarity of the generic $n$-dimensional Weyl-invariant quadratic curvature
gravity theories around their (anti-)de Sitter [(A)dS] and flat vacua. Weyl symmetry is spontaneously broken in
(A)dS and radiatively broken at the loop level in flat space. Save the three dimensional theory (which is
the Weyl-invariant extension of the new massive gravity), the graviton remains massless and the unitarity requires
that the only viable Weyl-invariant quadratic theory is the Weyl-invariant extension of the Einstein-Gauss-Bonnet
theory. The Weyl gauge field on the other hand becomes massive. Symmetry breaking scale fixes all the dimensionful
parameters in the theory.
\end{abstract}
\maketitle

\section{Introduction}
General Relativity (GR) is expected to be modified (or perhaps 
replaced by a new theory) at both large and small scales. For large distances one can keep the
theory intact with the high price of introducing a huge amount of
dark matter and dark energy to explain the flattening of the
galaxy rotation curves and the accelerated expansion of the
universe. In the UV regime, even though there is no
experimental necessity to modify GR, from the perturbative quantum
field theory perspective, it is a non-renormalizable theory with
no predictive power, hence it must be replaced by a UV-complete theory.

There are many proposals to modify GR at both of the regimes, here
we study the most general Weyl-invariant extension of  GR
augmented with the Weyl-invariant quadratic curvature terms built
from the contractions of the Riemann tensor. Thus we both extend
the diffeomorphism symmetry of GR to the local scale invariance
and add quadratic terms which give better UV behavior.
[Whether the theory really makes sense at the quantum level is an
open issue and there is not much work in the literature on the
quantization of Weyl-invariant gravity theories, see  \cite{shapiro}, and the earlier references therein, 
for discussions of the one-loop beta functions of the  Weyl-tensor square gravity and see {\cite{percacci} for a similar computation  in the conformally coupled scalar-tensor theory.] Local scale invariance
demands that the action has no dimensionful parameters, therefore
Newton's constant and any other dimensionful parameter, such as
the mass of graviton, the cosmological constant appears only after
the scale symmetry is broken (either by hand or spontaneously or
radiatively). [See \cite{tHooft} for a recent extension of the
Standard Model with local conformal symmetry.] For large distance
modifications of GR, as an alternative to the dark energy (or
cosmological constant), one could give a tiny mass to the graviton.
But in generic $n$ dimensions, there does not seem to be a
consistent way to give a non-linear mass to the graviton within
the frame of a single-metric theory. Ideally, one would like to
find a Lagrangian which includes gravity and a Higgs type scalar
field whose vacuum breaks the symmetry (not specified yet) and as
a result graviton becomes massive. If one considers the  rather
special case of the $n=2+1$ dimensions, such a theory exists: The
relevant symmetry is the Weyl symmetry [hence one not only has a
scalar field but also a non-compact Abelian gauge field and a
dynamical metric] and the action is given by the Weyl-invariant
extension \cite{DengizTekin, TanhayiDengizTekin} of the new
massive gravity (NMG) \cite{BHT1, BHT2}. NMG, currently, is the
only known parity invariant, non-linear extension of the
Fierz-Pauli massive gravity. This attractive feature of the model
led to a recent activity in the $2+1$ dimensional gravity theories
\cite{BHT1,BHT2,GulluTekin,deser,nakasone,liusun,canonical,cubic,clement1,giribet,clement2,gursesKilling,bakas,Aliev,grumiller,sinha,tahsin,ohta,naseh}.
Weyl-invariant extension of NMG \cite{DengizTekin,
TanhayiDengizTekin} goes one step further and provides (in AdS
not dS) an example of the mass generation for graviton
via the symmetry breaking mechanism in perfect analogy with the
Higgs mechanism in the Standard Model. For flat backgrounds, the
symmetry (Weyl symmetry) is broken at the two-loop level
\cite{tantekin} through the Coleman-Weinberg mechanism
\cite{coleman} and the graviton again gets a mass. All the
dimensionful scales in this theory are fixed by the symmetry
breaking scale.

In this paper we study the most general Weyl-invariant quadratic
gravity theory in $n$ dimensions with respect to its unitarity and
stability and explore the perturbative spectrum about the
(A)dS and flat backgrounds. Unitarity of the theory,
that is the ghost and tachyon freedom highly restricts the
spectrum and rules out a massive graviton for $n>3$, therefore no
viable extension of the Weyl-invariant NMG seems to exist beyond
$2+1$ dimensions. In fact, as we shall see, the only unitary
quadratic theory in generic $n$ dimensions is the Weyl-invariant
extension of the Einstein-Gauss-Bonnet theory.

The layout of the paper is as follows: In Section II, we briefly review the construction of the Weyl-invariant quadratic actions which was given
in \cite{DengizTekin}. In Section III, we find the maximally symmetric vacua of the theory and expand the action around the vacua up to quadratic order in the 
fluctuations of the fields. Section IV contains the discussion of the decoupling of the fields and in Section V, we find the masses of the free fields and explore the unitarity regions. Some computations which we used in finding the second order expansion of the action are summarized in the Appendix.  

\section{Weyl-invariant quadratic theory}

Generic $n$-dimensional Weyl-invariant quadratic theory was
constructed in \cite{DengizTekin} which we briefly recapitulate
here. [See also \cite{maki1,maki2} and   see \cite{oliva} for conformally invariant higher derivative scalar-tensor theories where the scalar field is coupled to the Euler densities  without the Weyl gauge field.] The literature on the Weyl
gauging (which was in fact the first example of upgrading a global
symmetry to a local one) of the metric is vast, for a quick review
see \cite{ORaif,ioro}. To grasp the bare essentials of the Weyl gauging
idea, consider the kinetic part of the scalar field action (with
the mostly plus signature):
\begin{equation}
S_{\Phi}=- \frac{1}{2}\int d^n x \sqrt{-g} g^{\mu \nu} \partial_\mu \Phi
\partial_\nu \Phi.
\end{equation}
For $n>2 $ dimensions, this action can be made locally
scale-invariant in a generally covariant way as follows \footnote{
$n=2$ case can also be handled similarly, but in this paper, we do
not consider this rather unique case.}
\begin{equation}\label{transform1}
 g_{\mu\nu} \rightarrow g^{'}_{\mu\nu}=e^{2 \zeta(x)} g_{\mu\nu}, \hskip 1 cm \Phi \rightarrow \Phi^{'} =e^{-\frac{(n-2)}{2}\zeta(x)} \Phi,
\end{equation}
where $\zeta(x) $ is a real function and the partial derivative should be replaced by the gauge covariant derivative as
\begin{equation}\label{covphi}
{\cal{D}}_\mu \Phi =\partial_\mu\Phi -\frac{n-2}{2} A_\mu \Phi,
\end{equation}
 with $ A_\mu $ being the non-compact Abelian gauge field transforming as
\begin{equation}\label{transform2}
 A_\mu \rightarrow A^{'}_\mu = A_\mu - \partial_\mu \zeta(x).
\end{equation}
Since we shall introduce dynamical gravity, we also need to know how the gauge covariant derivative acts on the metric. By simple inspection, one finds
\begin{equation}\label{transform3}
{\cal{D}}_\mu g_{\alpha \beta}=\partial_\mu g_{\alpha\beta}+
 2 A_\mu g_{\alpha \beta}.
\end{equation}
 By construction, these definitions yield
\begin{equation}
({\cal{D}}_\mu g_{\alpha \beta})^{'}=e^{2 \zeta(x)} {\cal{D}}_\mu g_{\alpha \beta},\hskip 0.5 cm
 ({\cal{D}}_\mu \Phi)^{'}=e^{-\frac{(n-2)}{2}\zeta(x)} {\cal{D}}_\mu \Phi,
\end{equation}
where (\ref{transform1}) and (\ref{transform2}) were employed. 
We also need  to write a kinetic term for the gauge field which comes with  a compensating scalar 
\begin{equation}
S_{A^\mu} =\varepsilon \int d^n x \sqrt{-g}\,\,
\Phi^{\frac{2(n-4)}{n-2}} F_{\mu\nu}F^{\mu\nu} ,
\label{maxwell}
\end{equation}
where $F_{\mu\nu}=\partial_\mu A_\nu-\partial_\nu A_\mu $ is the
usual gauge invariant field strength, and a dimensionless
parameter $\varepsilon$ is introduced (unitarity requirement will
later restrict it). Note also that for $ n =4 $, the
compensating factor drops out since the Maxwell action is already
conformally invariant. Weyl invariance in the gravity part follows
with the help of a Weyl-invariant "Christoffel connection" defined
as
\begin{equation}
 \widehat{\Gamma}^\lambda_{\mu\nu}=\frac{1}{2}g^{\lambda\sigma} \Big ({\cal{D}}_\mu g_{\sigma\nu}+{\cal{D}}_\nu g_{\mu\sigma}
-{\cal{D}}_\sigma g_{\mu\nu} \Big).
\label{christofel}
\end{equation}
From this connection a Weyl-invariant Riemann tensor can be defined as
 \begin{equation}
\widehat{R}^\mu{_{\nu\rho\sigma}} [g,A]=\partial_\rho
\widehat{\Gamma}^\mu_{\nu\sigma}-\partial_\sigma
\widehat{\Gamma}^\mu_{\nu\rho}
+ \widehat{\Gamma}^\mu_{\lambda\rho} \widehat{\Gamma}^\lambda_{\nu\sigma}-\widehat{\Gamma}^\mu_{\lambda\sigma} \widehat{\Gamma}^\lambda_{\nu\rho},
\end{equation}
which, after making use of (\ref{christofel}) becomes
 \begin{equation}
\widehat{R}^\mu{_{\nu\rho\sigma}} [g,A]
=R^\mu{_{\nu\rho\sigma}}+\delta^\mu{_\nu}F_{\rho\sigma}+2
\delta^\mu_{[\sigma} \nabla_{\rho]} A_\nu
+2 g_{\nu[\rho}\nabla_{\sigma]} A^\mu \\
 +2 A_{[\sigma} \delta_{\rho]}^\mu A_\nu +2 g_{\nu[\sigma}
A_{\rho]} A^\mu +2 g_{\nu[\rho} \delta_{\sigma]}^\mu  A^2,
\label{wiriem}
\end{equation}
where we have used the notation $2A_{[\rho}B_{\sigma]}\equiv
A_\rho B_\sigma-A_\sigma B_\rho$. Here  $\nabla_\mu$ is
the usual covariant derivative compatible with the Riemannian
metric $g_{ \mu \nu}$: For example one has $\nabla_\mu A_\nu =
\partial_\mu A_\nu - \Gamma_{\mu \nu}^\sigma A_\sigma$.  Note
that (\ref{wiriem}) does not have the original symmetries of the
Riemann tensor but this is not needed, all we want is to find a
short way to construct Weyl-invariant actions for generic higher
curvature gravity theories. For this purpose (\ref{wiriem}) can be
used. After contraction, Weyl-invariant Ricci tensor can be found
as
\begin{equation}
\begin{aligned}
\widehat{R}_{\nu\sigma} [g,A]&= \widehat{R}^\mu{_{\nu\mu\sigma}}[g,A] \\
&=R_{\nu\sigma}+F_{\nu\sigma}-(n-2)\Big [\nabla_\sigma A_\nu - A_\nu A_\sigma +A^2  g_{\nu\sigma} \Big ]-g_{\nu\sigma}\nabla \cdot A,
\label{wiricc}
\end{aligned}
\end{equation}
where $\nabla\cdot A\equiv \nabla_\mu A^\mu$. One more contraction
gives the scalar curvature
\begin{equation}
\widehat{R}[g,A]=R-2(n-1)\nabla \cdot A-(n-1)(n-2) A^2,
\label{wisclcr}
\end{equation}
which transforms like the inverse metric
as $(\widehat{R}[g,A])^{'} = e^{-2\zeta(x)}\widehat{R}[g,A]$. Weyl
invariance allows a potential term for the scalar field yielding
the action
\begin{equation}
S_\Phi=- \frac{1}{2}\int d^n x \sqrt{-g}\Big ({\cal{D}}_\mu
\Phi {\cal{D}}^\mu\Phi +\nu \,\Phi^{\frac{2n}{n-2}}\Big ) ,
\label{scalarwithpot}
\end{equation}
where $\nu \ge 0$ is a dimensionless coupling constant. Positivity
of $\nu$ is required for the existence of a ground state, further
constraints can come from the requirement of unitarity or the
existence of a maximally symmetric vacuum, as we shall discuss later. Collecting all
the pieces together, the generic Weyl-invariant quadratic action
is given as \cite{DengizTekin}
\begin{equation}
 S_{WI}= \int d^n x \sqrt{-g}\,\, \bigg\{\sigma\Phi^2\widehat{R}+\Phi^{\frac{2(n-4)}{n-2}}\Big[\alpha
 \widehat{R}^2
 +\beta \widehat{R}^2_{\mu\nu}+\gamma
 \widehat{R}^2_{\mu\nu\rho\sigma}\Big]\bigg\}+S_\Phi+S_{A^\mu},
\label{generic}
\end{equation}
where $S_{A^\mu}$ and  $S_\Phi$  are the Weyl-invariant actions
for gauge field and the scalar parts given by  (\ref{maxwell}) and
(\ref{scalarwithpot}). The action (\ref{generic}) is invariant
under the transformations (\ref{transform1}) and
(\ref{transform2}). Note that Weyl invariance alone leaves $7$
free dimensionless parameters, one of which can be eliminated by
scaling the total action. Here we have done the scaling in such a
way that the kinetic part of the scalar field action comes with the
canonical $ 1/2 $ factor.

Apparent simplicity of (\ref{generic}) is somewhat
deceptive since the explicit forms of the curvature terms yield a highly complicated action in terms of the
fields [$g_{\mu \nu}, A_\mu, \Phi $]. The explicit form of the scalar curvature square follows from (\ref{wisclcr})
\begin{equation}
\begin{aligned}
 \widehat{R}^2= &R^2-4(n-1)R(\nabla\cdot A)-2(n-1)(n-2)R A^2 \\
&+4(n-1)^2(\nabla \cdot A)^2 +4(n-1)^2(n-2)A^2(\nabla \cdot A) \\
& +(n-1)^2 (n-2)^2 A^4,
\end{aligned}
\label{wR}
\end{equation}
where $ \nabla.A=\nabla_\mu A^\mu $, $ A^2=A_\mu A^\mu $ and $ A^4=A_\mu A^\mu A_\nu A^\nu $. The square of the Ricci tensor follows from (\ref{wiricc})
\begin{equation}
\begin{aligned}
\widehat{R}^2_{\mu\nu}&= R^2_{\mu\nu}-2(n-2)R_{\mu\nu}\nabla^\nu
A^\mu-2R(\nabla\cdot
A)+2(n-2)R_{\mu\nu}A^\mu A^\nu-2(n-2)R A^2 \\
&-2(n-2)F^{\mu\nu}\nabla_\nu A_\mu
+F^2_{\mu\nu}+(n-2)^2(\nabla_\nu A_\mu)^2+(3n-4)(\nabla\cdot A)^2-2(n-2)^2A_\mu A_\nu\nabla^\mu A^\nu \\
&+(4n-6)(n-2)A^2(\nabla\cdot A) +(n-2)^2(n-1)(A)^4.
\end{aligned}
\label{wric}
\end{equation}
Finally, the square of the Riemann tensor follows from
(\ref{wiriem})
\begin{equation}
\begin{aligned}
\widehat{R}^2_{\mu\nu\rho\sigma}&=
R^2_{\mu\nu\rho\sigma}-8R_{\mu\nu}\nabla^\mu
A^\nu+8R_{\mu\nu}A^\mu A^\nu-
  4R A^2+nF^2_{\mu\nu}+4(n-2)(\nabla_\mu A_\nu)^2+4(\nabla\cdot A)^2\\
  &+8(n-2)(A)^2(\nabla\cdot A)-8(n-2)A_\mu A_\nu\nabla^\mu
  A^\nu+2(n-1)(n-2)(A)^4.
\end{aligned}
\label{wriem}
\end{equation}
Inserting (\ref{wR}), (\ref{wric}) and (\ref{wriem}) into
(\ref{generic}) yields an action in the Jordan frame with many
non-minimal interaction terms: For example the gauge field is
allowed to self interact at the  quadric  $A^4$ level, in a gauge
invariant way, even though this is an Abelian gauge theory.

\section{The action up to quadratic order in the fluctuations of the fields}

Now that we have constructed the most general Weyl-invariant
quadratic gravity theory, we can proceed to find its perturbative
spectrum, namely the particle content of the theory about its
maximally symmetric vacuum or vacua. Naively, one should find the
field equations first and find the constant curvature vacuum
solutions, then linearize the equations about any one of these vacua and
identify the propagating degrees of freedom. But, this is a very
complicated procedure, even the field equations are hard to find
(see \cite{ DengizTekin} for the field equations in the $n=3$
case). Instead of this cumbersome procedure, we will follow the
technique used in \cite{ubinmg} and directly expand the action
about its {\it{assumed}} maximally symmetric constant curvature
vacuum up to second order in the fluctuations of the fields. This
method not only will give the relation between the parameters of
the theory in the vacuum, but it also will simplify the
computation of the parts quadratic in the fluctuations of the
fields. For $n=3$, we have given the computation in
\cite{TanhayiDengizTekin}. Note that one does not need to go to the Einstein frame to determine the particle spectrum of the theory, 
one can directly work in the Jordan frame. Equivalence of the specific case of the conformally coupled scalar tensor theory with regard to its particle spectrum was shown in \cite{TanhayiDengizTekin}. 

Let us consider a dS or an AdS background (flat vacuum can be
obtained in the limit) for which the vacuum values of the fields
are
\begin{equation}
 \Phi_{vac}=m^{(n-2)/2}, \hskip 1 cm A^\mu_{vac}=0, \hskip 1 cm g_{\mu \nu}=\bar{g}_{\mu \nu},
\label{expectation}
\end{equation}
where $m$ is of the mass dimension and appears either by the
requirement that an (A)dS vacuum exists or as we shall briefly
discuss later, in the flat space case, conformal symmetry is
broken at the loop level via the Coleman-Weinberg mechanism
\cite{coleman,tantekin} where the scalar field acquires a non-zero
vacuum expectation value.  Hence in either case symmetry is broken in the
vacuum (\ref{expectation}). [There is a caveat here:
Strictly speaking, the Coleman-Weinberg potential was computed  in
three \cite{tantekin} and four dimensions \cite{coleman}, for larger dimensions we expect the symmetry to be broken at the loop level, but no explicit computation is available.] To
study the particle spectrum of the model we need to consider
fluctuations about the vacuum as
\begin{equation}
 \Phi=m^{(n-2)/2}+\tau \Phi_{L}, \hskip 1 cm   A_\mu=\tau A^{L}_\mu, \hskip 1 cm g_{\mu \nu}=\bar{g}_{\mu \nu}+\tau h_{\mu \nu},
\label{agg}
\end{equation}
where we have introduced $ \tau$, a small dimensionless parameter
to keep the track of the expansion orders, at the end it can be
set to unity. In what follows, we will use the conventions given
in \cite{ubinmg}. Let us represent the expansion of the action
(\ref{generic}) under the fluctuations (\ref{agg}) as follows:
 \begin{equation}
S_{WI}= \int d^n x \sqrt{-\bar{g}}\bigg \{{\cal L}(\tau^0)+\tau{\cal
L}(\tau^1)+\tau^2 {\cal L}(\tau^2)+\cdots\bigg \},
\end{equation}
where ${\cal L}(\tau^0)$ is a constant built from the vacuum values of the fields
which is irrelevant for the
perturbative spectrum that we shall discuss. On the other hand
${\cal L}(\tau^1)$ reads
\begin{equation}\label{firstorder}
{\cal L}(\tau^1) =  \Big (\frac{n}{n-2} m^{\frac{n-6}{2}} \Phi_L +
\frac{1}{4}  m^{n-4}h \Big )\Big({\cal C}\Lambda^2+4\sigma\Lambda
m^2-\nu m^4 \Big ),
\end{equation}
where
\begin{equation}
{\cal C}\equiv  \frac{8(n-4)}{(n-2)^2 } \Big ( n\alpha+\beta+\frac{2\gamma}{n-1} \Big ).
\end{equation}
Criticality of the action for arbitrary variations of the fields
about the vacuum, from  (\ref{firstorder}), yields
\begin{equation}
{\cal C}\Lambda^2+4\sigma\Lambda m^2-\nu m^4=0,
\label{vaceq}
\end{equation}
which reduces to the one given in
\cite{DengizTekin,TanhayiDengizTekin} for $n=3$ and for the NMG
condition ($\gamma =0$ and $ 8 \alpha + 3 \beta = 0$). One can
read (\ref{vaceq}) in two different ways: First, one can assume
that the symmetry is broken (namely $m$ is given), then $\Lambda$
is fixed by the symmetry breaking scale. Or, one can assume that $\Lambda$ is given (meaning that the existence of (A)dS vacuum is imposed) and $m$ is determined.  
Let us discuss the first case (the second case follows exactly the discussion in \cite{DengizTekin}). For $n \ne
4$, there are two vacua given as
\begin{equation}
\Lambda_\pm =  -\frac{ 2 m^2}{{\cal C}} \bigg [\sigma \mp \sqrt{
\sigma^2 + \frac{{\cal C}\nu }{4} }\, \bigg ], \hskip  1 cm  n\ne
4.
\end{equation}
As long as $\sigma^2 + \frac{{\cal C}\nu }{4}  \ge 0$, there is at least
one constant curvature vacuum. Four dimensions is unique in the sense that there is a
single vacuum
\begin{equation}
\Lambda = \frac{\nu m^2}{4 \sigma}.
\end{equation}
Finally, we need to find the  ${\cal L}(\tau^2)$ part of the
action, which requires a rather long computation.  A naive expansion of the quadratic parts of the action would result in a cumbersome expression which will not be explicitly background diffeomorphism invariant. To simplify the computation, one should use background diffeomorphism invariant quantities, such as the linearized Ricci scalar or the Einstein tensor as well as the self-adjointness of the involved operators and the linearized Bianchi identities. 
We give some essential parts of this computation in the Appendix and write
the result here
\begin{equation}
\begin{aligned}\label{quadgeneric}
 &{\cal L}(\tau^2)=  -\frac{1}{2}m^{n-4}h^{\mu\nu}\Big[\Big(\frac{4n}{n-2}\alpha+\frac{4}{n-1}\beta-\frac{8}{n-1}\gamma\Big)\Lambda
{\cal
G}^L_{\mu\nu}+(2\alpha+\beta+2\gamma)\Big(\bar{g}_{\mu\nu}\bar{\Box}-\bar{\nabla}_\mu\bar{\nabla}_\nu\Big)R_L\\
 &\hspace{2.4cm}+\frac{2\Lambda}{n-2}\Big(2\alpha+\frac{\beta}{n-1}-\frac{2(n-3)}{n-1}\gamma\Big)\bar{g}_{\mu\nu}R_L+(\beta+4\gamma)\bar{\Box}{\cal
G}^L_{\mu\nu}+\sigma m^2{\cal
G}^L_{\mu\nu}\Big]\\
&+m^{\frac{n-2}{2}}\Big[{\cal
C}\frac{\Lambda}{m^2}+2\sigma \Big] R_L
 \Phi_L+\frac{n}{2(n-2)}\Big[\frac{n(n-6){\cal
 C}}{(n-2)}\frac{\Lambda^2}{m^2}+4\sigma \Lambda-\frac{(n+2)}{n-2}m^2\nu\Big]\Phi_L^2 -\frac{1}{2}(\partial_\mu
 \Phi_L)^2\\
 &-m^{n-4}\Big[4(n-1)\alpha+n\beta+4\gamma\Big]R_L\bar{\nabla}\cdot
 A_L-m^{\frac{n-2}{2}}\Big[2(n-1){\cal
 C}\frac{\Lambda}{m^2}+4\sigma(n-1)+\frac{n-2}{2}\Big]\Phi_L\bar{\nabla}\cdot
 A_L\\
 &+m^{n-4}\Big[4(n-1)^2\alpha+n\beta+4\gamma\Big](\bar{\nabla}\cdot
 A_L)^2+\frac{1}{2}m^{n-4}\Big[(n^2-2n+2)\beta+2(3n-4)\gamma+2\varepsilon\Big](F_{\mu\nu}^L)^2\\
 &-2m^{n-2}\Big[\Big(2n(n-1)\alpha+(3n-4)\beta+8\gamma\Big)\frac{\Lambda}{m^2}+\frac{(n-1)(n-2)}{2}\sigma+\frac{(n-2)^2}{16}\Big]A_L^2,
\end{aligned}
\end{equation}
where $R^L$ and ${\cal G}_{\mu\nu}^L$ are the linearized
Ricci scalar and Einstein tensors defined by \cite{desertekin,desertekin2}:
\begin{equation}
\begin{aligned}
R^L=&\bar{\nabla}_\mu\bar{\nabla}_\nu
h^{\mu\nu}-\bar{\Box}h-\frac{2\Lambda}{n-2}h,\\
{\cal
G}_{\mu\nu}^L=&(R_{\mu\nu})_L-\frac{1}{2}\bar{g}_{\mu\nu}R^L-\frac{2\Lambda}{n-2}
h_{\mu\nu},\\
R_{\mu\nu}^L=&\frac{1}{2}\Big(\bar{\nabla}^\sigma\bar{\nabla}_\mu
h_{\sigma\nu}+\bar{\nabla}^\sigma\bar{\nabla}_\nu
h_{\sigma\mu}-\bar{\Box}h_{\mu\nu}-\bar{\nabla}_\mu\bar{\nabla}_\nu
h\Big).
\end{aligned}
\end{equation}
In (\ref{quadgeneric}), vacuum equation (\ref{vaceq}) has been
used. But, (\ref{quadgeneric}) is still a very complicated coupled
system of relativistic "harmonic oscillators" which need to be
decoupled before one can discuss what the true  propagating degrees of freedom
are. It pays to fix the Weyl gauge which
we shall do first below.

\section{Decoupling of the fields}
\subsection{Gauge fixing condition}
To remove the redundancy we need a proper gauge-fixing condition. Let the gauge-covariant derivative act on the gauge field as:
\begin{equation}
{\cal D}_\mu A_\nu\equiv \nabla_\mu A_\nu+(n-2) A_\mu A_\nu.
\end{equation}
Note that, compared to  (\ref{transform3}), we used $\nabla_\mu$
in the first term instead of the partial derivative. Under the
transformations (\ref{transform1}, \ref{transform2}), it is easy
to show that the divergence transforms as
\begin{equation}
(\mathcal{D}_\mu A^\mu)'=e^{-2\zeta(x)}\Big(\mathcal{D}_\mu
A^\mu-\mathcal{D}_\mu
\partial^\mu\zeta(x)\Big).
\end{equation}
Therefore, we can choose a Lorenz-like condition
\begin{equation}
\mathcal{D}_\mu A^\mu=\nabla\cdot A+(n-2)A^2=0.
\label{gaugefixing}
\end{equation}
 It is important to note that $\mathcal{D}_\mu \partial^\mu\zeta=0$ is also
Weyl-invariant. [This is a Weyl-invariant generalization of the
leftover gauge-invariance, $ \partial^2 \zeta=0 $, after the usual
Lorenz gauge $\partial_\mu A^\mu =0$ is chosen.] At the linear
level, (\ref{gaugefixing}) reduces to the background covariant
Lorenz gauge fixing condition : $\bar{\nabla}\cdot A_L=0$. With this choice, three terms in the action (\ref{quadgeneric}) drop out.

\subsection{Redefinition of the tensor field} In order to decouple the tensor and scalar fields, let us choose
\begin{equation}\label{generalrede}
h_{\mu\nu}=\widetilde{h}_{\mu\nu}-\frac{4}{n-2}m^{\frac{2-n}{2}}\bar{g}_{\mu\nu}\Phi_L,
\end{equation}
which yields the following relation between the linearized curvature expressions which we all need in the computation
\begin{equation}
\begin{aligned}
R_{\mu\nu}^L=&\widetilde{R}_{\mu\nu}^L+\frac{2}{n-2}m^{\frac{2-n}{2}}\Big((n-2)\bar{\nabla}_\mu\partial_\nu\Phi_L+\bar{g}_{\mu\nu}\bar{\Box}\Phi_L\Big),\\
 R_L=&\widetilde{R}_L+\frac{4}{n-2}m^{\frac{2-n}{2}}\Big((n-1)\bar{\Box}\Phi_L+\frac{2n}{n-2}\Lambda\Phi_L\Big),\\
{\cal G}_{\mu\nu}^L=&\widetilde{{\cal
G}}^L_{\mu\nu}+2m^{\frac{2-n}{2}}\Big(\bar{\nabla}_\mu\partial_\nu\Phi_L-\bar{g}_{\mu\nu}\bar{\Box}\Phi_L-2(n-2)\Lambda
\bar{g}_{\mu\nu}\Phi_L\Big),\\
h^{\mu\nu}{\cal G}^L_{\mu\nu}=&\widetilde{h}^{\mu\nu}{\cal
\widetilde{G}}^L_{\mu\nu}+4m^{2-n}\Big(m^{\frac{n-2}{2}}\widetilde{R}_L\Phi_L+2\frac{n-1}{n-2}\Phi_L\bar{\Box}\Phi_L+\frac{4n}{(n-2)^2}\Lambda\Phi_L^2\Big),\\
({\cal
G}_{\mu\nu}^L)^2=&({\cal{\widetilde{G}}}^L_{\mu\nu})^2+4m^{2-n}\Big((n-1)(\bar{\Box}\Phi_L)^2+\frac{4n}{(n-2)^2}\Lambda^2\Phi_L^2
+\frac{2(2n-1)}{n-2}\Lambda\Phi_L\bar{\Box}\Phi_L\Big)\\
&\hspace{3cm}+2m^{\frac{2-n}{2}}\Big((n-2)\widetilde{R}_L\bar{\Box}\Phi_L+2\Lambda\widetilde{R}_L\Phi_L\Big).
\end{aligned}
\end{equation}
After making use of  (\ref{generalrede}) and the gauge-fixing
condition and the vacuum equation (\ref{vaceq}) (which removes the
$\Phi_L^2$ term), the vector field is decoupled from the rest and
the quadratic part of (\ref{generic}) boils down to a more
transparent form
\begin{equation}
\begin{aligned}\label{quadgeneric3}
\widetilde{S}_{WI}= \int d^n x &\sqrt{-\bar{g}}\bigg
\{m^{n-4}\Big[-\Big(\frac{2n\Lambda}{n-2}\alpha+\frac{2\Lambda}{n-2}\beta-\frac{4(n-4)\Lambda}{(n-1)(n-2)}\gamma+\frac{m^2}{2}\sigma\Big)\widetilde{h}^{\mu\nu}
\widetilde{{\cal G}}^L_{\mu\nu}\\
&\hspace{1.1cm}+\Big(\alpha-\frac{n-4}{4}\beta-(n-3)\gamma\Big)\widetilde{R}_L^2+(\beta+4\gamma)
(\widetilde{{\cal
G}}^L_{\mu\nu})^2\Big]\\
&-\frac{1}{2}\Big[\frac{16}{(n-2)^2}\Big(2n(n-1)\alpha+(3n-4)\beta+8\gamma\Big)\frac{\Lambda}{m^2}+8\frac{(n-1)}{(n-2)}\sigma+1\Big](\partial_\mu
 \Phi_L)^2\\
 &+\frac{16}{m^2}\frac{(n-1)^2}{(n-2)^2}\Big(\alpha+\frac{n}{4(n-1)}\beta+\frac{1}{n-1}\gamma\Big)(\bar{\Box}\Phi_L)^2\\
 &+8m^{\frac{n-6}{2}}\frac{n-1}{n-2}\Big(\alpha+\frac{n}{4(n-1)}\beta+\frac{1}{n-1}\gamma\Big)\widetilde{R}_L\bar{\Box}\Phi_L\\
  &+\frac{1}{2}m^{n-4}\Big[(n^2-2n+2)\beta+2(3n-4)\gamma+2\varepsilon\Big](F_{\mu\nu}^L)^2\\
 &-2m^{n-2}\Big[\Big(2n(n-1)\alpha+(3n-4)\beta+8\gamma\Big)\frac{\Lambda}{m^2}+\frac{(n-1)(n-2)}{2}\sigma+\frac{(n-2)^2}{16}\Big]A_L^2\bigg\}.
\end{aligned}
\end{equation}
We still have the $\widetilde{R}_L\bar{\Box}\Phi_L$ coupling
between the tensor and the scalar fields. But this is not a problem
since it drops out once unitarity is imposed to remove the
higher-derivative Pais-Uhlenbeck term $(\bar{\Box}\Phi_L)^2$. It
is remarkable that both of these unwanted terms come with the same
coefficient, which must be set to zero:
\begin{equation}\label{condition}
\alpha+\frac{n}{4(n-1)}\beta+\frac{1}{n-1}\gamma=0.
\end{equation}
It is worth mentioning that  this condition is a necessary
but not  a sufficient condition for the unitarity of the theory.
 In three dimensions this condition gives the NMG theory after
noting that the Riemann tensor can be written as
$R^2_{\mu\nu\rho\sigma}=4R^2_{\mu\nu}-R^2$, which is the
Gauss-Bonnet identity, therefore it is easy to see that
(\ref{condition}) reduces to $8\tilde{\alpha}+3\tilde{\beta}=0$
once the square of the Riemann tensor is eliminated with the help
of the Gauss-Bonnet identity. Now that all the fields are
decoupled, we can find the masses and discuss the unitarity
regions in the theory.
\section{Particle spectrum and their masses}
In order to read the masses of the fields, let us write
(\ref{quadgeneric3}) as:
\begin{equation}
\label{quadgeneric4} \widetilde{S}_{WI}= \int d^n x
\sqrt{-\bar{g}} \bigg\{{\cal L}_{h_{\mu\nu}}+{\cal
L}_{A_\mu}+{\cal L}_\Phi\bigg\},
\end{equation}
where
\begin{equation}
\begin{aligned}
{\cal
L}_{h_{\mu\nu}}=&m^{n-4}\Big[-\Big(\frac{2n\Lambda}{n-2}\alpha+\frac{2\Lambda}{n-2}\beta-\frac{4(n-4)\Lambda}{(n-1)(n-2)}\gamma
+\frac{m^2}{2}\sigma\Big)\widetilde{h}^{\mu\nu}
\widetilde{{\cal G}}^L_{\mu\nu}\\
&\hspace{1.1cm}+\Big(\alpha-\frac{n-4}{4}\beta-(n-3)\gamma\Big)\widetilde{R}_L^2+(\beta+4\gamma)
(\widetilde{{\cal G}}^L_{\mu\nu})^2\Big],\\
{\cal L}_{A_\mu}=&\frac{1}{2}m^{n-4}\Big[(n^2-2n+2)\beta+2(3n-4)\gamma+2\varepsilon\Big](F_{\mu\nu}^L)^2\\
 &-2m^{n-2}\Big[\Big(2n(n-1)\alpha+(3n-4)\beta+8\gamma\Big)\frac{\Lambda}{m^2}+\frac{(n-1)(n-2)}{2}\sigma+\frac{(n-2)^2}{16}\Big]A_L^2,\\
{\cal L}_\Phi=&-\frac{1}{2}\Big[\frac{16}{(n-2)^2}\Big(2n(n-1)\alpha+(3n-4)\beta+8\gamma\Big)\frac{\Lambda}{m^2}+8\frac{(n-1)}{(n-2)}\sigma+1\Big](\partial_\mu
 \Phi_L)^2.
\end{aligned}
\end{equation}
The gauge field is of the massive Proca type and the scalar field
just has the kinetic term. On the other hand, the tensor part is
still complicated. Let us first look at the scalar and gauge field
parts.

One can choose $\varepsilon$ in such a way that the kinetic term
of the gauge field becomes zero which renders the gauge field
non-dynamical, but instead we choose
\begin{equation}
\varepsilon=-\frac{1}{2}\Big((n^2-2n+2)\beta+2(3n-4)\gamma+1/2\Big),
\end{equation}
to normalize the kinetic part to its canonical value $-\frac{1}{4}$ after scaling the field by a factor of $m^{\frac{(n-4)}{2}}$. We thus have
\begin{equation}
{\cal L}_{A_\mu}=-\frac{1}{4}(F_{\mu\nu}^L)^2
 -\frac{1}{2}M^2_A A_L^2,
\end{equation}
where the mass-square reads
\begin{equation}
M^2_A=4(n-4)\Big [2(n-1)\alpha+\beta \Big ] \Lambda+ \Big [ 2(n-1)(n-2)\sigma+\frac{(n-2)^2}{4}  \Big ] m^2 .
\label{massgauge}
\end{equation}
In getting this equation, we have imposed the constraint
(\ref{condition}) which was necessary for unitarity. Note that,
for the gauge field alone $M_A^2  \ge  0 $ is sufficient for
unitarity and in this case the theory has a massive (or massless
if the bound is saturated) spin-1 excitation. But, below we will
see that there will be another relation between $\alpha, \beta$
and $\gamma$ coming from the unitarity of the spin-2 part. In
three dimensions, (\ref{massgauge}) reduces to the one found in
\cite{TanhayiDengizTekin}.

On the other hand the scalar field part becomes
\begin{equation}
{\cal L}_\Phi = -\frac{ 4 M_A^2}{(n-2)^2 m^2 }\frac{1}{2}  (\partial_\mu \Phi_L)^2,
\end{equation}
which does not impose any new condition: The condition that makes
the gauge field non-tachyonic makes the scalar field non-ghost.
Also, when the gauge field has zero mass, the scalar field becomes
non-dynamical. For other values of $M_A^2$, the scalar field can be
re-scaled to have ${\cal L}_\Phi = -\frac{1}{2} (\partial_\mu
\Phi_L)^2$.

Finally, let us investigate the spin-2 part. The Lagrangian
density, following the procedure of \cite{BHT2, canonical}, can be
written in terms of two auxiliary fields $\varphi$ and
$f_{\mu\nu}$,  as
\begin{equation}
\begin{aligned}\label{auxil}
{\cal L}_{h_{\mu\nu}}=& a h^{\mu\nu}{\cal G}^L_{\mu\nu}(h)+b
R_L^2+c({\cal
G}^L_{\mu\nu}(h))^2\\
\equiv&a h^{\mu\nu}{\cal G}^L_{\mu\nu}(h)+f^{\mu\nu}{\cal
G}^L_{\mu\nu}(h)+\varphi
R_L-\frac{m_1^2}{2}\varphi^2-\frac{m_2^2}{4}(f_{\mu\nu}^2-f^2),
\end{aligned}
\end{equation}
where  $f \equiv \bar{g}^{\mu \nu} f_{ \mu \nu}$ and
\begin{equation}
\begin{aligned}
a\equiv &
-m^{n-4}\Big(\frac{2n\Lambda}{n-2}\alpha+\frac{2\Lambda}{n-2}\beta-\frac{4(n-4)\Lambda}{(n-1)(n-2)}\gamma+\frac{m^2}{2}\sigma\Big),\\
b\equiv &
m^{n-4}\Big(\alpha-\frac{n-4}{4}\beta-(n-3)\gamma\Big),\\
c\equiv & m^{n-4}(\beta+4\gamma).
\end{aligned}
\end{equation}
We have to first determine $m_i^2$ in terms of the parameters of
the theory. This can be done by using the field equations for the
auxiliary fields
\begin{equation}
\begin{aligned}
f_{\mu\nu}=&\frac{2}{m_2^2}{\cal
G}^L_{\mu\nu}(h)+\bar{g}_{\mu\nu}\frac{n-2}{(n-1)m_2^2}R_L, \hskip
1.2 cm \varphi=&\frac{1}{m_1^2}R_L,
\end{aligned}
\end{equation}
in (\ref{auxil}) to obtain
\begin{equation}
{\cal L}_{h_{\mu\nu}}=a h^{\mu\nu}{\cal
G}^L_{\mu\nu}(h)+\frac{1}{m_2^2}({\cal
G}^L_{\mu\nu}(h))^2+\Big(\frac{1}{2m_1^2}-\frac{(n-2)^2}{4(n-1)m_2^2}\Big)R_L^2,
\end{equation}
which yields
\begin{equation}
c=\frac{1}{m_2^2},\,\,\,\,\mbox{and}\,\,\,\,\,
b=\frac{1}{2m_1^2}-\frac{(n-2)^2}{4(n-1)m_2^2}.
\end{equation}
Then (\ref{auxil}) reads as
\begin{equation}\label{elif}
\begin{aligned}
{\cal L}_{h_{\mu\nu}}=(a h^{\mu\nu}+f^{\mu\nu}){\cal
G}^L_{\mu\nu}(h)+\varphi
R_L-\frac{\varphi^2}{4b+c(n-2)^2/(n-1)}-\frac{1}{4c}(f_{\mu\nu}^2-f^2).
\end{aligned}
\end{equation}
The unitarity condition (\ref{condition}) in this parametrization becomes
\begin{equation}\label{phi-condition}
4b+c\frac{(n-2)^2}{n-1}=0,
\end{equation}which decouples  $\varphi$  and drops the term $\varphi R_L$ in (\ref{elif}). Therefore, the gravity part becomes:
\begin{equation}
\label{gravitypart}{\cal L}_{h_{\mu\nu}}=(a
h^{\mu\nu}+f^{\mu\nu}){\cal
G}_{\mu\nu}^L(h)-\frac{1}{4c}(f_{\mu\nu}^2-f^2).\end{equation}
 In order to obtain  the spectrum and read the masses, one should decouple the
$h_{\mu\nu}$ and $f_{\mu\nu}$ fields which can be done with the
following redefinition (assuming $ a \ne 0$, see below for the
$a=0$ case)
\begin{equation}h_{\mu\nu}=\texttt{h}_{\mu\nu}-\frac{1}{2a}f_{\mu\nu}.\end{equation}
Inserting this  in (\ref{gravitypart}) results in:
\begin{equation}
\begin{aligned}\label{gravity-de}
{\cal L}_{h_{\mu\nu}}=a \texttt{h}^{\mu\nu}{\cal
G}_{\mu\nu}^L(\texttt{h})-\frac{1}{4a}f^{\mu\nu}{\cal
G}_{\mu\nu}(f)-\frac{1}{4c}(f^2_{\mu\nu}-f^2).
\end{aligned}
\end{equation}
The first term is just like the linearized part of the pure
(cosmological) Einstein-Hilbert theory with an effective Newton's
constant, therefore it propagates a massless spin-2 field as long
as $ a < 0$. The second part is the Lagrangian density of a
massive Fierz-Pauli spin-2 field. Because of the coefficient of
the kinetic part, it is clear that massless and massive spin-2
fields cannot be unitary at the same time. The only solution is to
freeze the massive spin-2 field by giving it an infinite mass:
$c=0 $, which, together with our earlier unitarity condition
(\ref{condition}), yield
 \begin{equation}
4\gamma+\beta=0,\hspace{1cm} \alpha=\gamma.
\end{equation}
This is exactly the Einstein-Gauss-Bonnet theory which has a massless unitary excitation in (A)dS as long as \cite{Sisman}
\begin{equation}\label{cond_on_sigma}
\sigma>-\frac{4(n-3)(n-4)\gamma\Lambda}{(n-1)(n-2)m^2}.
\end{equation}
Therefore in  $n \ge 4 $ dimensions, out of all the Weyl-invariant
quadratic gravity theories, unitarity condition singled out the
Weyl-invariant extension of the Einstein-Gauss-Bonnet theory. The
mass of the gauge field in this theory reduces to
\begin{equation}
M_A^2=8(n-3)(n-4)\gamma
\Lambda+\Big(\frac{(n-2)^2}{4}+2(n-1)(n-2)\sigma\Big)m^2.
\end{equation}
Note that $M_A^2 \ge 0 $, a  condition for the unitarity of the
scalar and gauge field parts, gives
\begin{equation} \sigma\geq
-\frac{4(n-3)(n-4)\gamma\Lambda}{(n-1)(n-2)m^2}-\frac{n-2}{8(n-1)},
\end{equation}
which is a weaker condition than  (\ref{cond_on_sigma}). Finally
one can check that the stronger unitarity condition
(\ref{cond_on_sigma}), which now becomes $\sigma  > - \frac{{\cal
C} \Lambda}{2}$, is compatible with the existence of a maximally
symmetric vacuum (\ref{vaceq}) for both dS and AdS spaces.

Consider now the critical case, $a=0$ in (\ref{gravitypart}) for which one has
\begin{equation}
\label{gravitypart2}{\cal L}_{h_{\mu\nu}}= h^{\mu\nu}{\cal
G}_{\mu\nu}^L(f)-\frac{1}{4c}(f_{\mu\nu}^2-f^2),\end{equation}
where we have used the self-adjointness of the involved operators
in the first term. Now variation with respect to $h_{\mu \nu}$
yields
\begin{equation}
{\cal G}_{\mu\nu}^L(f) =0,
\end{equation}
 which can be solved as \cite{BHT2}
 \begin{equation}
f_{\mu\nu}=\bar{\nabla}_\mu B_\nu+\bar{\nabla}_\nu B_\mu.
\end{equation}
Inserting this into (\ref{gravitypart2}) gives up to boundary terms
\begin{equation}\label{critical}
{\cal L}_{h_{\mu\nu}}=-\frac{1}{4c}F_{\mu\nu}^2
-\frac{2\Lambda}{c(n-2)} B_\mu^2,
\end{equation}
where $ F_{\mu\nu}$ is the field strength of the $B$-field.
For $c >0$, (\ref{critical}) describes a massive spin-1 excitation
with $M^2 = \frac{ 4 \Lambda}{n-2}$ in analogy with the critical
point of NMG \cite{BHT2}. One can check that the unitarity
conditions found before are compatible with the criticality
condition and the full Weyl-invariant quadratic theory propagates
a unitary massive Weyl gauge field and a massless scalar field in
addition to the just discussed massive spin-1 field in dS. 

\subsection{Flat background for $n=4$}
In the flat backgrounds in four dimensions, as we noted before, the
symmetry is broken via the Coleman-Weinberg mechanism
\cite{coleman} for which the one-loop effective potential
becomes $V(\Phi) =c_1\Phi^4 ( \log (\Phi/m) + c_2)$, where the
actual values of the constants are not relevant here: All that
matters is that the scalar field gets a vacuum expectation value
and the symmetry gets broken. Then (\ref{condition}) together with
$\beta+4\gamma=0$ gives
\begin{equation}
\begin{aligned}
{\cal
L}_{h_{\mu\nu}}=&-\frac{m^2}{2}\sigma\widetilde{h}^{\mu\nu}
\widetilde{{\cal G}}^L_{\mu\nu},\\
{\cal
L}_{A_\mu}=&-\frac{1}{4} (F^L_{\mu\nu})^2-\frac{1}{2}\Big (1+12 \sigma \Big) m^2 A_L^2,\\
{\cal
L}_\Phi=&-\frac{1}{2}\Big(1+12\sigma\Big)(\partial_\mu
 \Phi_L)^2.
\end{aligned}
\end{equation}
So in the Weyl-invariant extension of the Einstein-Gauss-Bonnet
theory in four dimensions there is   a unitary massless spin-2 field,
a massive spin-1 and a massless spin-0 field for $\sigma >0$.
Newton's constant and the mass of the gauge field are fixed by the
vacuum expectation value of the scalar field. [Note that any other $n>4$ theory has the same spectrum.]

\section{Conclusions}

We have found the perturbative particle spectrum and discussed the unitarity of the $n$-dimensional Weyl-invariant quadratic gravity, constructed in \cite{DengizTekin}, about their (A)dS and flat backgrounds. Three dimensional case, that is the Weyl-invariant extension of the new massive gravity, studied at length in \cite{TanhayiDengizTekin}, has the remarkable property that through the breaking of the symmetry, the graviton gets a unitary Fierz-Pauli type mass. The hope was to extend this mass-generation mechanism to four and more dimensions, but, as we have shown in this paper, the only unitary theory, beyond three dimensions, among the  Weyl-invariant quadratic theories is the Weyl-invariant Einstein-Gauss-Bonnet model which propagates a massless spin-2 particle as well as  massive spin-1 and  massless spin-0 particles. [ For an interesting Higgs-type mechanism in the first order formalism of gravity see  \cite{percacci2}.]
Now that we have shown the unitarity of the theory, it would be interesting to study the Weyl-invariant Einstein-Gauss-Bonnet theory in four dimensions with respect to its black hole and cosmological solutions.

\section{appendix}
We would like to expound upon our computation of the second order action in the fluctuations of the fields here.  The expressions below are valid up to  ${\cal O}(\tau^2)$.
The scalar field action reads
\begin{equation}
\begin{aligned}
S_\Phi=-\frac{1}{2}\int
d^nx\sqrt{-\bar{g}}&\bigg\{ \nu m^n+\tau  \Big[\frac{m^n}{2}h+2m^{\frac{n+2}{2}}\frac{n}{n-2}\Phi_L\Big] \nu \\
&+\tau^2\Big[(\partial_\mu\Phi_L)^2+(n-2)m^{\frac{n-2}{2}}\bar{\nabla}\cdot
A_L \Phi_L+\frac{(n-2)^2}{4}m^{n-2}A_L^2\\
&+\frac{n(n+2)}{(n-2)^2}m^2 \nu \Phi_L^2+\frac{n}{n-2}m^{\frac{n+2}{2}} \nu h\Phi_L+\frac{m^n}{8} \nu h^2-\frac{m^n}{4}  \nu h_{\mu\nu}^2\bigg\}.
\end{aligned}
\end{equation}
The coupling term between the scalar field and the Ricci scalar yields up to quadratic order 
\begin{equation}
\begin{aligned}
\int d^nx\sqrt{-g}\Phi^2R =\\
\int
d^nx\sqrt{-\bar{g}}m^{n-2}&\bigg\{\frac{2n}{n-2}\Lambda+\tau\Big[\frac{n}{n-2}\Lambda
h+R_L+\frac{4n}{n-2}\Lambda m^{\frac{2-n}{2}}\Big]\\
&-\frac{1}{2}\tau^2\Big[h^{\mu\nu}(R_{\mu\nu})_L+\frac{1}{2}h
R_L+\frac{n-4}{n-2}\Lambda h^2_{\mu\nu}-\frac{1}{2}\Lambda h^2-h
R_L-\frac{4n}{n-2}\Lambda m^{2-n}\Phi_L\Big]\bigg\}.
\end{aligned}
\end{equation}
To expand the quadratic curvature parts of the action, the following steps proved useful:
First let us write
\begin{equation}
\begin{aligned}
&\int d^nx\sqrt{-g}\Phi^{\frac{2(n-4)}{n-2}}\Big(\alpha R^2+\beta R_{\mu\nu}^2+\gamma R_{\mu\nu\rho\sigma}^2\Big)=\\
&\int d^nx\sqrt{-\bar{g}}\bigg\{\Big(m^{\frac{n-2}{2}}+\tau
\Phi_L\Big)^{\frac{2(n-4)}{n-2}}\Big(\bar{X}+\tau X^{(1)}+\tau^2
X^{(2)}\Big)\bigg\},
\end{aligned}
\end{equation}
 where one finds
\begin{equation}
\begin{aligned}
\bar{X} \equiv&\frac{n{\cal C}}{2(n-4)}\Lambda^2,\\
X^{(1)} \equiv &\frac{n{\cal C}}{4(n-4)}\Lambda^2 h+\frac{(n-2){\cal
C}}{2(n-4)}\Lambda R_L.
\end{aligned}
\end{equation}
Now in order to find $X^{(2)}$, we rewrite it as:
\begin{equation}
\begin{aligned}
X^{(2)}&=\Big[\sqrt{-g}\Big(\alpha R^2+\beta R_{\mu\nu}^2+\gamma
R_{\mu\nu\rho\sigma}^2\Big)\Big]^{(2)}\\
=&\Big[\sqrt{-g}\Big((\alpha-\gamma) R^2+(\beta+4\gamma)
R_{\mu\nu}^2+\gamma  \chi_E \Big)\Big]^{(2)},
\end{aligned}
\end{equation}
where $\chi_E  \equiv R_{\mu\nu\rho\sigma}^2-4R_{\mu\nu}^2+R^2$ is the Gauss-Bonnet combination. From
\cite{desertekin,desertekin2}, one can write the $X^{(2)}$ as:
\begin{equation}
\begin{aligned}
X^{(2)}=-\frac{1}{2}h^{\mu\nu}\Big[&\Big(\frac{4n\Lambda}{n-2}\alpha+\frac{4\Lambda}{n-1}\beta-\frac{8\Lambda}{n-1}\gamma\Big){\cal
G}_{\mu\nu}^L\\
&+(2\alpha+\beta+2\gamma)\Big(\bar{g}_{\mu\nu}\bar{\Box}-\bar{\nabla}_\mu\bar{\nabla}_\nu\Big)R_L+\frac{2\Lambda}{n-2}\Big(2\alpha+
\frac{1}{n-2}\beta-\frac{2(n-3)}{n-1}\gamma\Big)\bar{g}_{\mu\nu}R_L\\
&+(\beta+4\gamma)\bar{\Box}{\cal G}_{\mu\nu}^L+\frac{{\cal
C}}{4}\Lambda^2 h_{\mu\nu}-\frac{{\cal
C}}{8}\Lambda^2\bar{g}_{\mu\nu}h\Big].
\end{aligned}
\end{equation}

\section{\label{ackno} Acknowledgments}

We would like to thank Tahsin C. Sisman for useful discussions.
The work of  B.T. is supported by the TUBITAK Grant No. 110T339.
S.D. is supported by  TUBITAK Grant No. 109T748.


\begin{thebibliography}{0}

\bibitem{shapiro} 
  G.~de Berredo-Peixoto and I.~L.~Shapiro,
  ``Conformal quantum gravity with the Gauss-Bonnet term,''
  Phys.\ Rev.\ D {\bf 70}, 044024 (2004).

\bibitem{percacci}
  R.~Percacci,
  ``Renormalization group flow of Weyl invariant dilaton gravity,''
  New J.\ Phys.\  {\bf 13}, 125013 (2011).

\bibitem{tHooft}
  G.~'t Hooft,
 ``A class of elementary particle models without any adjustable real parameters,''
  Found.\ Phys.\  {\bf 41}, 1829 (2011).

\bibitem{DengizTekin} S.~Dengiz and B.~Tekin, ``Higgs Mechanism for New Massive Gravity and Weyl Invariant Extensions of Higher Derivative Theories,''
Phys.\ Rev.\ D\ \textbf{84}, 024033  (2011).

\bibitem{TanhayiDengizTekin} M.~R.~Tanhayi, S.~Dengiz and B.~Tekin, ``Unitarity of Weyl-Invariant New Massive Gravity and
Generation of Graviton Mass via Symmetry Breaking,''
arXiv:1112.2338, To appear in Phys. Rev. D.

\bibitem{BHT1} E.~A.~Bergshoeff, O.~Hohm and P.~K.~Townsend, ``Massive Gravity in Three Dimensions,''
Phys.\ Rev.\ Lett.\ \textbf{102}, 201301 (2009).

\bibitem{BHT2} E.~A.~Bergshoeff, O.~Hohm and P.~K.~Townsend, ``More on Massive 3D Gravity,'' Phys.\ Rev.\  D {\bf 79}, 124042 (2009).

\bibitem{GulluTekin} I.~Gullu and B.~Tekin, ``Massive Higher Derivative Gravity in D-dimensional Anti-de Sitter Spacetimes,'' Phys.\ Rev.\ D \textbf{80},
064033 (2009).

\bibitem{deser} S.~Deser, ``Ghost-free, finite, fourth order D=3 (alas) gravity,'' Phys.\ Rev.\ Lett. \textbf{103}, 101302
(2009).

\bibitem{nakasone} M.~Nakasone and I.~Oda, ``On Unitarity of Massive Gravity in Three Dimensions,'' Prog.\ Theor.\ Phys.
\textbf{121}, 1389 (2009).

\bibitem{liusun} Y.~Liu and Y.~W.~Sun, ``On the Generalized Massive Gravity in $AdS_3$,'' Phys.\ Rev.\ D \textbf{79},
126001 (2009).

\bibitem{canonical} I.~Gullu, T.~C.~Sisman and B.~Tekin, ``Canonical Structure of Higher Derivative Gravity in
3D,'' Phys.\ Rev.\ D \textbf{81}, 104017 (2010).

\bibitem{cubic} I.~Gullu, T.~C.~Sisman and B.~Tekin, ``All Bulk and Boundary Unitary Cubic Curvature Theories in Three Dimensions,''
Phys.\ Rev.\ D \textbf{83}, 024033 (2011).

\bibitem{clement1} G.~Clement, ``Warped AdS(3) black holes
in new massive gravity,'' Class.\ Quant.\ Grav.\ \textbf{26}, 105015
(2009).

\bibitem{giribet} E.~Ayon-Beato, G.~Giribet and M.~Hassaine, {}``Bending
AdS Waves with New Massive Gravity,'' JHEP \textbf{0905}, 029 (2009).

\bibitem{clement2} G.~Clement, ``Black holes with a
null Killing vector in new massive gravity in three dimensions,''
Class.\ Quant.\ Grav.\ \textbf{26}, 165002 (2009).

\bibitem{gursesKilling} M.~Gurses, ``Killing Vector Fields in Three
Dimensions: A Method to Solve Massive Gravity Field Equations,'' Class.\ Quant.\ Grav.\ \textbf{27},
205018 (2010).

\bibitem{bakas} I.~Bakas, C.~Sourdis, 
``Homogeneous vacua of (generalized) new massive gravity,'' Class.\ Quant.\ Grav.\ \textbf{28},
015012 (2011).

\bibitem{Aliev} H.~Ahmedov and A.~N.~Aliev, ``Exact Solutions
in D-3 New Massive Gravity,'' Phys.\ Rev.\ Lett.\ \textbf{106},
021301 (2011).

\bibitem{grumiller}
  D.~Grumiller and O.~Hohm,
  ``$AdS_3/LCFT_2$ - Correlators in New Massive Gravity,''
  Phys.\ Lett.\  B {\bf 686}, 264 (2010).

\bibitem{sinha}
  A.~Sinha,
  ``On the new massive gravity and AdS/CFT,''
  JHEP {\bf 1006}, 061 (2010).

\bibitem{tahsin} 
  I.~Gullu, T.~C.~Sisman and B.~Tekin,
  ``Born-Infeld extension of new massive gravity,''
  Class.\ Quant.\ Grav.\  {\bf 27}, 162001 (2010).

\bibitem{ohta}
  N.~Ohta,
  ``A Complete Classification of Higher Derivative Gravity in 3D and
  Criticality in 4D,''
  Class.\ Quant.\ Grav.\  {\bf 29}, 015002 (2012).

\bibitem{naseh} 
  M.~Alishahiha and A.~Naseh,
  ``Holographic renormalization of new massive gravity,''
  Phys.\ Rev.\ D {\bf 82}, 104043 (2010).

\bibitem{tantekin} P.~N.~Tan, B.~Tekin and Y.~Hosotani,
``Spontaneous Symmetry Breaking at Two Loop in 3-d Massless Scalar
Electrodynamics,'' Phys.\ Lett.\  B\ \textbf{388}, 611 (1996);
``Maxwell-Chern-Simons Scalar Electrodynamics at Two Loop,''
Nucl.\ Phys.\  B\ \textbf{502}, 483 (1997).

\bibitem{coleman}
  S.~R.~Coleman and E.~J.~Weinberg,
  ``Radiative Corrections as the Origin of Spontaneous Symmetry Breaking,''
  Phys.\ Rev.\ D {\bf 7}, 1888 (1973).

\bibitem{maki1}
  T.~Maki, Y.~Norimoto and K.~Shiraishi,
  ``On the cosmology of Weyl's gauge invariant gravity,''
  Acta Phys.\ Polon.\  B {\bf 41} 1195 (2010).

\bibitem{maki2}
  T.~Maki, N.~Kan, K.~Kobayashi and K.~Shiraishi,
  ``Flux vacua in DBI type Einstein-Maxwell theory,''
  arXiv:1109.4687 [gr-qc].

\bibitem{oliva} 
  J.~Oliva and S.~Ray,
  ``Conformal couplings of a scalar field to higher curvature terms,''
  arXiv:1112.4112 [gr-qc].

\bibitem{ORaif} L.~O'Raifeartaigh, I. Sachs and C. Wiesendanger, Meeting on 70 Years of Quantum
Mechanics, Calcutta, India, 1996, edited by P. Bandyopadhyay.

\bibitem{ioro} 
  A.~Iorio, L.~O'Raifeartaigh, I.~Sachs and C.~Wiesendanger,
  ``Weyl gauging and conformal invariance,''
  Nucl.\ Phys.\ B {\bf 495}, 433 (1997).

\bibitem{Sisman}
  T.~C.~Sisman, I.~Gullu and B.~Tekin,
  ``All unitary cubic curvature gravities in D dimensions,''
  Class.\ Quant.\ Grav.\  {\bf 28}, 195004 (2011).


\bibitem{ubinmg} I.~Gullu, T.~C.~Sisman and B.~Tekin, ``Unitarity analysis of general Born-Infeld gravity
theories,'' Phys.\ Rev.\ D\ \textbf{82}, 124023  (2010).

\bibitem{desertekin} S. Deser, B. Tekin, ``Energy in Generic Higher Curvature Gravity Theories,'' Phys. Rev. D 67,  084009
(2003).

\bibitem{desertekin2}
  S.~Deser and B.~Tekin,
  ``Gravitational energy in quadratic curvature gravities,''
  Phys.\ Rev.\ Lett.\  {\bf 89}, 101101 (2002).

\bibitem{percacci2} 
  R.~Percacci,
  ``The Higgs phenomenon in quantum gravity,''
  Nucl.\ Phys.\ B {\bf 353}, 271 (1991).

\end{thebibliography}
\end{document}